\documentstyle[aps]{revtex}

\begin{document}
\title{Dark Energy and the Nature of the Graviton}
\author{A. Zee}
\address{Kavli Institute for Theoretical Physics\\
University of California\\
Santa Barbara, CA 93106\\
USA\\
zee@kitp.ucsb.edu}
\maketitle

\begin{abstract}
Does the existence of dark energy suggest that there is more to the graviton
than we think we know?
\end{abstract}

\bigskip

\bigskip The word paradox has been emasculated by indiscriminate usage in
the physics literature. A real paradox should involve a major and clear-cut
discrepancy between theoretical expectation and experimental measurement.
The ultraviolet catastrophe for example is a paradox, the resolution of
which around the dawn of the 20th century ushered in quantum physics.
Surely, the most egregious paradox of physics around the dawn of the 21st
century is the cosmological constant paradox\cite{weinberg}.

The root of the paradox lies in a fundamental clash between Einstein's view
and the particle theorist's view of gravity. To particle physicists, the
graviton is just another particle, or a particular mode of the vibrating
string. Indeed, given that a massless spin 2 particle couples to the
stress-energy tensor, one can reconstruct Einstein's theory. According to
Einstein, however, gravity has to do with the curvature of spacetime, the
arena in which all fields and particles live in. The graviton is not just
another particle.

The graviton is not just another particle --- it knows too much. The
electromagnetic force knows about the particles carrying charge, and the
strong force knows about the particles carrying color. But the gravitational
force knows about anything carrying energy and momentum, including an
apparently innocuous constant shift in the Lagrangian density ${\cal L}%
\rightarrow {\cal L}-\Lambda $.

As is well known, the paradox can be easily described. The natural value of $%
\Lambda $ in particle physics is expected by dimensional analysis to be $\mu
^{4}=\mu /(\mu ^{-1})^{3}$ for some relevant mass scale $\mu $ where the
second form of writing $\mu ^{4}$ reminds us that $\Lambda $ is a mass or
energy density$.$ Whether one associates $\mu $ with grand unification,
electroweak symmetry breaking, or the quark confinement transition and
consequently has a value of order $10^{19}$ $Gev,$ $10^{2}$ $Gev,$ and $1$ $%
Gev$ respectively is immaterial. The natural value $\Lambda \sim \mu
^{4}=\mu /(\mu ^{-1})^{3}$ is outrageous even if we take the smallest value
for $\mu .$ We don't even have to put in actual numbers to see that there is
a humongous discrepancy between theoretical expectation and observational
reality. We know the universe is not permeated with a mass density of order
of $1$ $Gev$ on every cube of size $1$ $(Gev)^{-1}.$

The cosmological constant paradox is basically an enormous mismatch between
the units natural to particle physics and natural to cosmology. Measured in
units of $Gev^{4}$ the cosmological constant is so incredibly tiny that
particle physicists have traditionally assumed that it must be
mathematically zero, and have looked in vain for a plausible mechanism to
drive it to zero. One of the disappointments of string theory is its
inability to resolve the cosmological constant paradox.

But Nature has a big surprise for us. While theorists racked their brains
trying to come up with a convincing argument that $\Lambda =0,$
observational cosmologists\cite{riess}\cite{perl} steadily refined their
measurements and changed their upper bound to an approximate equality\cite
{neutr} 
\begin{equation}
\Lambda \sim (10^{-3}ev)^{4}!!!
\end{equation}

The cosmological constant paradox deepens. Theoretically, it is easier to
explain why some quantity is mathematically $0$ than why it happens to be $%
\sim 10^{-124}$ in the units natural (if indeed they are) to the problem.

As is also well known, strictly speaking, we should refer to the observation
of the cosmological constant as the observation of a hiherto unknown dark
energy since we do not know the equation of state associated with the
observed energy density $(10^{-3}ev)^{4}.$

To make things worse, the energy density $(10^{-3}ev)^{4}$ happens to be the
same order of magnitude as the present matter density of the universe $\rho
_{M}.$ This is sometimes refer as the cosmic coincidence problem. The
cosmological constant $\Lambda $ is, within our traditional understanding, a
parameter in the Lagrangian. On the other hand, since most of the mass
density of the universe resides in the rest mass of baryons, as the universe
expands $\rho _{M}(t)$ decreases like $(1/R(t))^{3}$ where $R(t)$ denotes
the scale size of the universe. In the far past, $\rho _{M}$ was much larger
than $\Lambda ,$ and in the far future, it will be much smaller. It just so
happens that in this particular epoch of the universe, when we are around,
that $\rho _{M}\sim \Lambda .$ Or to be less anthropocentric, the epoch when 
$\rho _{M}\sim \Lambda $ happens to be when galaxy formation has been
largely completed. In their desperation, some theorists have even been
driven to invoke anthropic selection\cite{banks}\cite{weinberg2}\cite
{vilenkin}.

My impression is that theoretical physicsts outside the high energy
community are not completely aware of how desperate the situation is, but
people are grasping at straws and a number of outlandish suggestions have
been aired. In this spirit I would like to offer a thought I have
entertained for some time but did not ``dare'' to publish. When I wrote my
recent field theory textbook I sketched\cite{zee} what I had in mind in
passing. It may be worthwhile to elaborate on what I wrote there and to
bring it to the attention of a broader audience.

In the development of physics, there have been numerous instances of
reasoning by historical precedent or analogy. For example, when confronted
with data showing that the energy of the electron in an atom is quantized
physicists recalled that the vibrational frequency of a violin string is
also quantized. As we all know, this turns out to be an apt analogy as both
the energy of the electron and the vibrational frequency of a string are
given by the eigenvalues of linear partial differential equations. I suggest
that perhaps similarly we can ask if historically there have been cases of a
physical quantity initially thought to be 0 but then turned out to be
extremely small but not precisely 0. I suspect that the proton decay rate
may be an apt example and that it may shed some light on the cosmological
constant paradox.

\bigskip Let us go through the story of proton decay. To make my point I
will take some liberty with history. Suppose that in 1953 some theorists
were to calculate the rate $\Gamma $ for protons to decay in the natural
mode $p\rightarrow e^{+}+\pi ^{0}.$ The interaction of the pion with the
proton and the neutron was known to be described by a term like $g\pi 
\overline{n}p$ in the Lagrangian with $g$ a dimensionless coupling of order
1. These theorists would naturally construct a Lagrangian out of the
available fields, namely the proton field $p,$ the electron field $e,$ and
the pion field $\pi ,$ and thus write down something like $f\pi \overline{e}p
$ with some constant $f.$ Note that $\pi \overline{e}p$ has mass dimension 4
and hence $f$ is dimensionless just like $g$. Since $\pi \overline{e}p$
violates isospin invariance, the theorists would expect $f$ to be suppressed
relative to $g$ by some measure of isospin breaking, say the fine structure
constant $\alpha .$ The natural value for $\Gamma $ would then come out to
be many many orders of magnitude larger than the experimental upper bound on 
$\Gamma .$ The theorists would then set $\Gamma =0$ and cast about for an
explanation. After an enormous struggle, the theorists were unable to come
up with a compelling explanation and this failure became known as the proton
decay rate paradox.

Eventually, someone with great authority and prestige in the community,
namely Wigner, decreed the law of baryon number conservation. Surely, even
in the unthinkably primitive days of 1953 this would have been recognized as
a pronouncement and not as an explanation. (The pronouncement could be
dressed up formally by imposing a $U(1)$ transformation under which $%
p\rightarrow e^{i\theta }p$ while $e$ and $\pi $ do not change and requiring
that the Lagrangian remains invariant.) But there would have been no deep
understanding of this astonishing discrepancy between theoretical
expectation and experimental upper bound.

Indeed, imagine an alternative history in which, while other important
particle physics experiments were being performed in 1957, some intrepid
experimentalist, ignoring conventional theoretical wisdom, actually went out
and measured the proton decay rate to be some tiny but non-zero value. The
proton decay rate paradox would have deepened, much as how the cosmological
constant paradox deepened with the discovery of a tiny $\Lambda .$

Let us now review how the proton decay rate paradox was resolved
historically. The first remark is that the eventual explanation did not
emerge within the orthodox theory fashionable in 1957, nor did it come from
an understanding of some kind of mechanism causing protons to decay, but
rather it came totally from ``left field'', from a study of baryon
spectroscopy, which led to the notion of quarks. The correct degrees of
freedom are not given by the proton and pion fields $p$ and $\pi ,$ but by
the quark fields $q.$ The effective Lagrangian ${\cal L}$ is to be
constructed out of quark $q$ and lepton $l$ fields and must satisfy the
symmetries that we know. Three quarks disappear, so we write down
schematically $qqq,$ but three spinors do not a Lorentz scalar make. We have
to include a lepton field and write $qqql.$ Since four fermion fields are
involved, these terms have mass dimension 6 and so in ${\cal L}$ they have
to appear as $\frac{1}{M^{2}}qqql$ with some mass $M,$ corresponding to the
mass scale of the physics responsible for proton decay. Thus, the
probability of proton decay is proportional to ($\frac{1}{M^{2}})^{2}=\frac{1%
}{M^{4}}.$ By dimensional reasoning, we obtain the proton decay rate $\Gamma
\sim (\frac{m_{p}}{M})^{4}m_{p}.$ The absurdly small value of $\Gamma $ is
then naturally explained by the fourth power of the small number $(\frac{%
m_{p}}{M})$ for $M$ big enough. No mystery left!

Note that in principle all of this could be done as soon as Gell-Mann
introduced the notion of quarks in 1964, long before anybody even dreamed of
a grand unified theory with proton decay.

As long we are discussing revisionist, but possible, history, we can imagine
some brilliant theorist in another civilization far far away puzzling over
the proton decay rate paradox eventually realizing that the key to
explaining an absurdly small number is to promote the dimension of the
effective Lagrangian merely from 4 to 6. In hindsight, we can say that the
extremely long lifetime of the proton could have pointed to the existence of
quarks.

I would like to raise the question whether the cosmological constant paradox
might not be solved in the same way. Perhaps the gravitational field $g_{\mu
\nu }$ is the analog of the proton and pion field $p$ and $\pi .$ The high
energy and more fundamental degrees of freedom in the gravitational field
may not be the metric $g_{\mu \nu },$ but some mysterious analog of the
quark field $q.$ This may emerge as a construct in string or M theory, or it
could be something else completely. In the history of the proton decay
paradox as recounted by me, there is an additional twist, namely that the
degree of freedom $q$ is confined and not physical. Before the advent of
quantum chromodynamics, theorists could only write $p\sim qqq,$ without any
clear idea about what the symbol $\sim $ might mean. We are in a similar
position here: the metric $g_{\mu \nu }$ might be a composite object, but I
certainly do not know what it is a composite of, and the objects of which $%
g_{\mu \nu }$ is a composite may also be as observable or as unobservable as
the quarks.

The cosmological term $\Lambda \sqrt{g}$ in the Lagrangian has mass
dimension 0 and we somehow have to promote 0 to a higher number. One
difficulty with this view is of course how we could possibly promote the
dimension of the cosmological term without at the same time changing the
mass dimension of the Einstein-Hilbert term $\frac{1}{G}\sqrt{g}R.$ Our
historical analogy may again be helpful: the 1953 view that the pion nucleon
coupling term has dimension $4$ turns out to be correct. While the dimension
4 term $\pi \overline{e}p$ was replaced by the dimension 6 term $qqql$ the
dimension 4 term $\pi \overline{n}p$ was replaced by dimension 4 terms of
the form $\overline{q}Aq$ with $A$ a gluon potential. The dimension of one
of the terms gets promoted while the dimension of the other term remains the
same. So it is entirely conceivable to me that the cosmological constant
term could end up with a higher dimension while the Einstein-Hilbert term
either remains dimension 4 or is replaced by dimension 4 terms. Thus,
suppose the cosmological constant term actually has dimension $p>0$ so that
it is given in the Lagrangian by a term of the form $\frac{1}{M^{p-4}}{\cal O%
}$ with $M$ some mass scale characteristic of the deeper structure of the
graviton, perhaps the same as the Planck mass, perhaps not. The observed
cosmological constant would then be given by $\Lambda \sim \frac{1}{M^{p-4}}<%
{\cal O}>=(\frac{m}{M})^{p}M^{4},$ where the expectation value of the
operator ${\cal O}$ in the physical universe $<{\cal O}>=m^{p}$ is set by
physics at some low energy scale $m.$ With $m$ small enough, and or $p$ big
enough, we could easily get the suppression factor we want.

As hinted above, I even suspect that the Lagrangian formalism, being a
mathematical realization of the quasi-theological (at least historically)
variational principle, may well be wrong. The cause of all our trouble is
that the flat space Lagrangian ${\cal L}$ could always be shifted by a
constant ${\cal L}\rightarrow {\cal L}-\Lambda $ without changing its
variation. But this also points to the mystery of quantum mechanics\cite
{feynmanquest} because without quantum mechanics we could have lived happily
with equations of motion without ever bothering with the Lagrangian.

Another possible way to nullify the physical consequence of the shift ${\cal %
L}\rightarrow {\cal L}-\Lambda $ is to postulate that $g=\det g_{\mu \nu }$
is not a dynamical variable. This was proposed\cite{w+z} 20 years ago as an
explanation of why $\Lambda $ is mathematically zero. But with $\Lambda $
now known to be tiny but non-zero this avenue seems to me less promising.

I also could not resist mentioning another wild speculation\cite{gravitipole}%
. Many years ago, inspired by the almost exact correspondence between
Einstein's post-Newtonian equations of gravity and Maxwell's equations of
motion I proposed the gravitipole in analogy with Dirac's magnetic monopole.
After Dirac there was considerable debate on how a field theory of magnetic
monopoles may be formulated. Eventually, 't Hooft and Polyakov showed that
the magnetic monopole exists as an extended solution in certain non-abelian
gauge theories. Most theorists now believe that electromagnetism is merely a
piece of a grand unified theory and that magnetic monopoles exist. Might it
not turn out that Einstein's theory is but a piece of a bigger theory and
that gravitipoles exist? In grand unified theory the electromagnetic field
is a component of a multiplet. Could it be that the gravitational field also
somehow carries an internal index and that the field we observe is just a
component of a multiplet? Throwing caution to the wind, I also asked in\cite
{gravitipole} if the gravitipole and the graviton might not form a
representation under some dual group just as the magnetic monopole and the
photon form a triplet under the dual group of Montonen and Olive\cite{mo}.

Perhaps we do not know as much about the graviton as we think we do.

\bigskip

{\bf {\large Acknowledgments\medskip }}

This work was supported in part by the National Science Foundation under
grant number PHY 99-07949.


\begin{references}
\bibitem{feynmanquest}  When I discussed the speculative idea proposed in 
\cite{w+z} with R. Feynman, he asked me if I knew of a formalism in quantum
mechanics in which one could calculate the difference between two energy
levels in say an atom directly (which is after all what experimentalists
measure) without having to calculate the two energy levels separately and
then subtract one from the other. Without gravity, only differences in
energy matter.
\end{references}
\end{document}